# A 350mK, 9T scanning tunneling microscope for the study of superconducting thin films on insulating substrates and single crystals

Anand Kamlapure, Garima Saraswat, Somesh Chandra Ganguli, Vivas Bagwe and Pratap Raychaudhuri[*]

*Department of Condensed Matter Physics and Materials Science, Tata Institute of Fundamental Research, Homi Bhabha Rd., Colaba, Mumbai 400 005, India.*

and

Subash P. Pai

*Excel Instruments, 28, Sarvodaya Industrial Premises, Off Mahakali Caves Road, Andheri (East), Mumbai 400 093, India.*

We report the construction and performance of a low temperature, high field scanning tunneling microscope (STM) operating down to 350mK and in magnetic fields up to 9T, with thin film deposition and in-situ single crystal cleaving capabilities. The main focus lies on the simple design of STM head and a sample holder design that allows us to get spectroscopic data on superconducting thin films grown *in-situ* on insulating substrates. Other design details on sample transport, sample preparation chamber and vibration isolation schemes are also described. We demonstrate the capability of our instrument through the atomic resolution imaging and spectroscopy on NbSe$_2$ single crystal and spectroscopic maps obtained on homogeneously disordered NbN thin film.

[*] E-mail: pratap@tifr.res.in

## I. Introduction

Ever since its invention, the scanning tunneling microscope (STM)[1,2] has emerged as a powerful spectroscopic tool, combining atomic scale imaging with simultaneous spectroscopic capabilities. One particularly attractive feature of STS is its unsurpassed energy resolution ( <100 µeV ), limited by thermal broadening of the Fermi edge. This makes the study of unconventional superconductors using STS particularly attractive since the characteristic energy scales, given by the superconducting energy gap, $\Delta$, vary from a fraction of a meV to few meV. However, the low transition temperatures of these superconductors also require the STM to operate at temperatures well below 1K. Precaution needs to be taken that both tip and sample are at thermal equilibrium at the same temperature during the measurement.

Over the past few years there are a few[3,4,5,6,7,8,9,10,11] STM which can reach temperatures below 1K incorporating in-situ cleaving and surface cleaning of single crystals and thin film deposition capabilities. However, a design that allows measurements to be performed on in-situ grown superconducting films on insulating substrates has remained a challenge. The study of superconducting thin films on insulating substrates is important for several reasons. First, single crystalline substrates of insulating MgO, $SrTiO_3$ and $LaAlO_3$ remain the most popular choice for growing lattice matched high-quality epitaxial thin films of several superconductors used for basic studies and applications, such as $YBa_2Cu_3O_7$, $(La,Sr)_2CuO_4$ and NbN. More importantly, the superconducting transition temperature of a superconductor in contact with a normal metal is suppressed through proximity effect up to a depth of the order of the coherence length, $\xi$, from the interface. Since $\xi$ (few nanometers to a few hundred nanometers), is in the same range of the thickness as most epitaxial films it is important to use insulating substrates in order to study the intrinsic properties of superconducting films.

In this paper, we describe the construction of low temperature STM (LT-STM) with base temperature of 350mK specifically designed for spectroscopic investigations of *in-situ* superconducting thin films grown on insulating substrates in a deposition chamber connected to the STM. In addition, our design also incorporates a crystal cleaving assembly for the study of superconducting single crystals. The highlights of our STM are a simple stable design of STM head and a molybdenum sample holder which allows deposition of superconducting thin films on insulating substrates up to a deposition temperature of $800^0$C. While most of our measurements are restricted below 12 K the temperature of the LT-STM can be precisely controlled from 350 mK – 20 K with temperature drift < 10 mK below 3K and < 20 mK in the range 3 - 20 K over 8 hours. We demonstrate topographic imaging with atomic resolution and spectroscopic imaging down to 350 mK with and without magnetic field.

**II. Setup**

The overall schematic of our system is shown in Fig. 1. The LT-STM assembly consists of three primary sub-units: (i) The sample preparation chamber, (ii) the load lock chamber to transfer the sample from the deposition chamber to the STM and (iii) the $^4$He dewar with 9T magnet housing $^3$He cryostat on which the STM head is attached. The $^4$He dewar hangs from a specially designed vibration isolation table mounted on pneumatic legs. A combination of active and passive vibration isolation systems are used to obtain the required mechanical stability of the tip. Data acquisition is done using the commercial R9 SPM controller from RHK Technology[12]. In following subsections we describe the mechanical details of various components of the setup.

**II.A. STM head**

Over the years, several designs of STM heads have been adopted for operation at low temperatures[13,14,15] based on the requirement of stability and convenience of sample or tip exchange. Some of the popular designs include the Pan type[3,16] and Besocke Beetle-type[17,18], which involve coordinated control of multiple piezo elements for coarse positioning. In contrast, the design of our STM head, which is directly mounted below the $^3$He pot (Fig. 2) is relatively simple. In this design both coarse approach as well as scanning is achieved through movement of the tip whereas the sample is static. The outer body is made of single piece of gold plated oxygen free high conductivity (OHFC) Copper. The sample holder, coming from the top with the sample facing down, engages on a Gold plated Copper part which is electrically isolated from the main body using cylindrical Macor[19] machinable ceramic part. Both these parts are glued together using commercially available low temperature glue[20]. The copper part has 45° conical cut at the top matching with sample holder. In the conical region, there are two nonmagnetic stainless steel studs where sample holder gets locked and it can be disengaged from vertical manipulator. The copper part also has two leaf springs made of phosphor bronze which grab the sample holder and also provide better thermal contact and prevent mechanical vibration of the sample holder. Electrical contact to this copper part is given by soldering a stud which extrude from the lower side. Positioning unit is located in the cuboidal cavity in the lower part of STM head. One of the sides of the cavity is open to get access for mounting the positioning unit and changing tip. The positioning unit consists of a coarse approach positioner and a piezoelectric tube on which the STM tip is fixed. To bring the tip within tunneling range of the sample we use a coarse positioner[21] which works on the principle of slip-stick motion. The coarse positioner is fixed to a copper bottom plate using a pair of titanium screws which are in turn screwed to the main body. Fine positioning and scanning is performed using a 1 inch long piezoelectric tube[22] which has gold plated electrodes inside and outside. Outside gold plating is divided into two segments. The lower half is used for Z motion while upper segment has four quadrants and used for XY motion. Inner electrode is grounded and wrapped out on the upper side to avoid the build-up of any static charge. The piezo-tube is electrically isolated from coarse positioner at the bottom and the copper tip carrier on the top through Macor[19] pieces which are glued to the tube so as to reduce differential thermal expansion. The copper tip holder is glued on the upper side of top Macor piece. We use Pt-Ir wire (80-20%) of diameter 300µm as tip which is held frictionally in 400µm bore that is drilled on tip holder. Tip is prepared by cutting the Pt-Ir wire using a sharp scissor at an angle and subsequently field emitted in vacuum at low temperature to achieve the desired sharpness. Printed circuit boards screwed on the three sides of the cuboid serve as the connecting stage for electrical connection to the piezo units, sample and tip. Temperature of STM head is measured using two Cernox™ sensors[23] mounted on the bottom plate of the STM as well as on the $^3$He pot. The entire STM head is enclosed in gold plated copper can ensuring temperature homogeneity over the entire length of the head. We observe that after achieving a stable temperature for about 10 min the temperature of the STM head and $^3$He pot differ at most by 20mK. The piezo-constants for the scanning head were initially calibrated using lithographically patterned Au lines of width 100 nm and separation of 100 nm on a metallic substrate, and subsequently the same calibration was verified through atomic resolution topographic image on NbSe$_2$.

**II. B. Sample holder**

The main challenge in the study of superconducting films grown in-situ on insulating substrates is in establishing the electrical contact with the sample for doing STM experiments. We overcome this problem by using a design of a sample holder where the film can be directly grown in-situ on the insulating substrate fixed on the holder, and subsequently transferred to STM head for measurement. The sample holder made of molybdenum is shown in Fig. 3. The choice of the material is given by a trade-off between the need of high thermal conductivity to ensure temperature homogeneity during measurements and the capacity to withstand temperature up to $800^0$C during deposition in reactive atmosphere (e.g. oxygen and nitrogen). The substrate is mounted with silver epoxy on the top flat surface and fixed in position by fastening a cap having 4.3 mm diameter hole in the center. The edge of the cap makes direct contact with the top surface of the sample and brings it in electrical contact with the rest of the sample holder. The lower part of the sample holder ends in a 45° slant which mates with the corresponding part on the STM head as shown in Fig. 2. The sample holder has a horizontal M4 threads on the side for mounting on the horizontal manipulator and M6 threads at the bottom for mounting it on the vertical manipulator. It has two diametrically opposite cuts at the bottom side which fits on the studs on STM head and locks the circular movement while disengaging the sample holder from the vertical manipulator after the sample is mounted on the STM head.

For STM measurements on films grown on insulating substrates, first two contact strips are deposited ex-situ on two edges of the substrate as shown in Fig. 3(b). The width of the strips is adjusted such that when the substrate with contact pads is mounted on the sample holder, a small portion of the strip on either side is exposed through the hole in the cap (Fig. 3(d)). When the superconducting film is deposited on the substrate in the in-situ chamber, the edge of the film is in contact with the strip and is therefore electrically connected to the entire sample holder. In principle, the strips could be made of any material that can withstand the deposition temperature of the superconducting film. However, in most cases we found it convenient to make the strips of the same material as the material under study. Since our STM head is symmetric, the tip engages at the center of the sample which ensures tip to strip distance ~1mm. This is much larger than the length over which we would expect superconducting proximity effects from the contact pads to play any role in the measurements.

For the study of single crystals a single piece sample holder of similar shape without the cap is used. The crystal is mounted on the flat surface using a two component conducting silver epoxy[24]. Depending on the hardness, the crystal is cleaved in vacuum (in the load lock cross) alternatively by gluing a small rod on the surface using the same silver epoxy and hitting it with a hammer or by gluing a tape on the surface and pulling the tape using one of the manipulators.

**II.C. Sample preparation chamber**

The sample preparation chamber, fitted with a turbo molecular pump and with a base pressure ~ $1\times10^{-7}$ mbar, is located on the top of table and is connected to load lock through a gate valve (Fig 4). The chamber consists of two magnetron sputtering guns facing down at an angle, to the substrate heater. The confocal arrangement of guns allows for co-sputtering. The substrate heater consists of a resistive heating element made of a patterned molybdenum plate. Sample holder is inserted using the horizontal manipulator in the chamber through the load-lock and held above the heater. It is heated radiatively and its temperature is measured using thermo-couple (PT100) located inside, at the tip of the horizontal manipulator. In addition, the chamber also contains a plasma ion source[25] for cleaning substrates prior to deposition and two tungsten boats for thermal evaporation.

**II. D. Load lock and sample manipulators**

The load-lock, located at the top of the $^3$He cryostat, has six CF35 ports and it is connected to sample preparation chamber and STM chamber through gate valves. Typical time to pump the load-lock chamber from ambient pressure to $1\times10^{-6}$ mbar is about 20 minutes. Sample manipulators (Fig. 5) are made of seamless steel tubes (closed at one end) and have matching threads at the end to engage on the corresponding threads on the sample holder. A thermocouple is fitted inside the horizontal sample manipulator to measure the temperature of the sample during deposition. Once the sample is deposited horizontal manipulator is pulled back bringing the sample holder in the cross, and the sample holder is transferred to vertical manipulator and inserted into the STM head.

**II. E. Cryostat and temperature control of the sample**

The low temperature stage consists of an internally fitted charcoal sorption pump based $^3$He cryostat from Janis Research Company[26] (Fig 6). We use a custom design with annular shaped sorption pump, 1K pot and $^3$He pot which give us direct line of site access from the top of the cryostat to the STM head mounted below the $^3$He pot. To ensure thermal stability the STM head is bolted to $^3$He pot using 6 screws which ensures good thermal contact between the two. To prevent radiative heating, a radiation plug is inserted in the cryostat after loading the sample using the same vertical manipulator as the one used to insert the sample. The radiation plug (not shown) sits just above the STM head. The $^3$He pot and sorption pump are fitted with resistive heaters. All electrical wires coming from the top are thermally anchored at the 1K pot and the $^3$He pot. The entire process of cooling the STM from 4.2 K to the base temperature of 350 mK takes about 20 min with a hold time of about 8 hrs. We wait for about 15 min for after the base temperature is reached before starting our measurements. Between the base temperature and 3 K, we control the temperature by controlling the temperature of the sorption pump. For stabilizing above 3K we use the resistive heater fitted on the $^3$He pot by keeping the sorption pump temperature constant at 30 K.

**II.F. Liquid Helium Dewar**

The cryostat is mounted in a 65 liters capacity Al-Fibreglass Dewar with retention time of approximately 5 days. The superconducting magnet with maximum of 9 T aligned along the STM tip hangs from the top flange of the cryostat. Exhaust line of the cryostat is connected with one way valve which maintains a constant pressure slightly above atmosphere. This allows us to flow liquid $^4$He in a capillary wrapped around the sorption pump such that the sorption pump can be cooled without using an external pump.

**II.G. Vibrational and electrical noise reduction**.

Most crucial part of any STM design is the vibrational and electrical noise reduction as it is directly reflected in the ultimate noise level in the tunneling current. We have adopted three isolation schemes to reduce vibrational noise. For sound isolation, the entire setup is located in a sound proof enclosure made of sound proofing perforated foam. To reduce vibrational noise mainly coming from the building, the entire setup rests on a commercial vibration isolation table[27] (Newport SmartTable®) with integrated active and passive stages with horizontal and vertical resonant frequency < 1.7 Hz. Finally, since in our cryostat the 1K pot pump has to be on during STM operation, special precaution has to be taken to isolate the system from the pump vibrations which get transmitted in two different ways: (i) Direct pump vibration transmitted through vibration of the connecting bellows and (ii) indirect vibration transmitted through the sound propagated through the $^4$He gas in the pumping line. The first source is isolated by keeping the pumps on a different floor in the basement and a rigid section of the pumping line is embedded in a heavy concrete block before connecting to the pump. To isolate the second source of vibration a special pumping scheme is adopted. The 1K pot is connected to the pump through two alternate pumping lines. While condensing the $^3$He and cooling the STM head from 4.2K to the base temperature, the 1K pot is cooled to 1.6 K by pumping through a 25.4 mm diameter pumping line directly connected to the pump. Once the base temperature of 350 mK is reached on the STM head, the first pumping line is closed and the second pumping line is opened. This line has a 30 cm long 10 cm diameter intermediate section packed with high density polystyrene foam which isolates the STM from the sound generated by the pump. Since the polystyrene foam reduces the pumping speed, the 1K pot warms up to 2.8 K, with no noticeable increase in the temperature of the STM head. During the steady-state operation of the STM at 350 mK the pumping is further reduced by partially closing a valve to keep the 1K pot at a constant temperature of ~ 4 K. While operating in this mode we do not observe any difference in vibration level on the top of the cryostat with the 1K pot pump on or off as shown in Fig. 7(a).

To reduce the electrical noise coming from the 50Hz line signal, ground connection of all instruments, table and Dewar are made to a separate master ground. RF noise is further reduced by introducing 10 MHz low pass filter before each connection that goes into the STM. The tunneling current is detected using a Femto DLPCA-200 current amplifier placed at the top if the cryostat with gain of $10^9$ V/A. While the bandwidth of the DLCPA-200 amplifier is 500 kHz, the measurement bandwidth is set digitally restricted to 2.5 kHz in the R9 SPM controller.

The final test of isolation performance is obtained from the spectral density (SD) in the current and Z-height signals. We recorded these signals at 350 mK in actual operating condition. Figure 7(b) shows the SD of the current (i) when the tip is out of tunneling range (background noise of to electronics), (ii) at a fixed tunneling current with feedback on condition, and (iii) after switching off the feedback for 5 s. The SD with tip out of tunneling range is below 300 fA Hz$^{-1/2}$. At fixed tunneling current (feedback on) additional peaks appear in the SD at 25.5 Hz and 91.5 Hz but the peak signal is only marginally larger than 300 fA Hz$^{-1/2}$. Even after switching off the feedback the peak signal is less than 1 pA Hz$^{-1/2}$. Similarly, the Z-height SD at fixed tunneling current with feedback on (Fig. 7(c)) is less than 2 pm Hz$^{-1/2}$ at all frequencies and less than 50 fm Hz$^{-1/2}$ above 150 Hz. The low Z-height and current noise allows us to get very good signal to noise ratio in spectroscopic measurements which we will show in the next section.

## III. STM performance

In this section we demonstrate the spatial and energy resolution of our LT-STM through measurements on Pb single crystal, NbSe$_2$ single crystal and NbN thin films.

### III. A. Tunneling spectroscopy on Pb

To characterize the energy resolution of our system the tunneling spectrum was acquired on a Pb single crystal. For spectroscopy *dI/dV* vs. *V* is measured with feedback switched off using standard modulation technique using an internal lock-in built-in within our main control unit (R9, RHK technology). Fig. 8 shows the typical spectrum acquired at 500mK at a single point on polished Pb single crystal using modulation voltage of 150µV and frequency 419.3 Hz. We have verified that the tunneling spectra do not show any noticeable variation when modulation frequency is varied from 400 Hz to 2.5 kHz. Fig. 8 also shows the fit using Bardeen-Cooper-Schrieffer theory for tunneling conductance which is described by the tunneling equation given by[28],

$$G(V) = \frac{dI}{dV}|_V = \frac{d}{dV}\left\{\frac{1}{R_N}\int_{-\infty}^{\infty} N_S(E)[f(E) - f(E-eV)]\,dE\right\}$$

where $N_S(E) = Re\{\frac{|E|}{\sqrt{(|E|)^2 - \Delta^2}}\}$ . We have taken into account broadening due to the finite modulation voltage which is used for lock-in measurements by doing adjacent averaging of points in the theoretical curve over a sliding voltage range of 150 µV. BCS fit gives an energy gap Δ=1.3 meV which is in good agreement[29] with reported values of the energy gap in Pb.

**III.B. Atomic resolution and vortex imaging on NbSe$_2$**

To test our system for atomic resolution and in magnetic field, we performed measurements on a 2H-NbSe$_2$ single crystal. Having a hexagonal closed packed layered structure this crystal can be easily cleaved in-plane. We cleaved the crystal in-situ by attaching a tape on the surface and subsequently pulling the tape in vacuum in the load-lock chamber using the sample manipulators. Fig. 9 shows the atomic resolution image at 350mK which reveals the hexagonal lattice structure. The lattice spacing of 0.34 nm is in very good agreement with the lattice constant of NbSe$_2$ known from literature.[9, 30,31]

For imaging the vortex state, we have first taken full area spectroscopic map over an area of 352 × 352 nm in magnetic field of 200mT at 350mK. In this measurement we recorded the spatially resolved tunneling spectra (*dI/dV vs. V*) at each point of a grid having 64 × 64 pixels by sweeping the bias from 6mV to -6mV. Figure 10(a-d) shows intensity plots of the tunneling conductance normalized at 6mV at different bias voltages, showing the hexagonal vortex lattice. The lattice constant, $a \approx 109.8$ nm is in excellent agreement with the theoretical value expected from Ginzburg Landau theory[28,32]. For voltages below Δ/*e* the vortices appear as regions with larger conductance whereas for voltages close to the coherence peak the vortices appear as regions with lower conductance. Fig. 10(e) shows the line scan sectioned on the line shown in Fig. 10(a). Three representative spectra are highlighted in the figure. Spectra 1 and 3 correspond to the superconducting region while the spectrum 2 is at the vortex core and has a zero bias conductance peak which is the signature of Andreev bound state inside the vortex core. In figure 10(f) we show a high resolution (128 × 128) conductance map obtained by measuring *dI/dV* at a fixed bias voltage of 1.4mV while scanning over the same area.

**III.C. Scanning tunneling spectroscopy on disordered NbN thin film**

As an example of a measurement on thin films we show spatially resolved STS measurements on a disordered epitaxial NbN thin film grown on single crystalline MgO substrate. This kind of sample is expected to show variations in the local superconducting order parameter which in turn reflects in the spatial variation of the coherence peak heights and the zero bias conductance. The sample was grown in-situ by sputtering Nb in the Ar+$N_2$ mixture keeping the susbtrate holder[33] at $600^0$C. $T_c$ of the sample measured by four probe technique (after completing the STS measurements) is 6.4K. For spatially resolved spectroscopy 6 tunneling conductance spectra were acquired at each point on a 32×32 grid over an area of 200×200nm at 500 mK. Figure 11 (a)-(b) show the average of 6 spectra acquired at each point along a 200 nm line and the average of all the spectra over the entire area. Both individual and average spectra show the presence of a superconducting gap and coherence peaks at the gap edge. The coherence peaks are partially suppressed compared to the expectation from BCS theory which is a hallmark of a strongly disordered superconductor[34].

**IV. Summary**

In summary, we have presented a relatively simple design of a $^3$He refrigerator based STM operating down to 350 mK and in magnetic fields up to 9 T. Using a novel design for sample holder and STM head we demonstrate the capability to perform STM and STS measurements on in-situ cleaved single crystals and in-situ grown thin films.

**Acknowledgements**

We would like to thank Supriya Dhawde for preparing the 3D drawing of the experimental systems, Parasharam Shirage and A Thamizhavel for providing the NbSe$_2$ single crystals, John Jesudasan for providing the optimal conditions for deposition of NbN thin films, and Sourin Mukhopadhyay for being involved in the fabrication of an early $^4$He version of the STM. The work was supported by Department of Atomic Energy, Government of India.

[16] Chr. Wittneven, R. Dombrowski, S. H. Pan, and R. Wiesendanger, Rev. Sci. Instrum. **68**, 3806 (1997).

[17] K. Besocke, Surface Sci. **181,** 145 (1987).

[18] J. Frohn, J. F. Wolf, K. Besocke, and M. Teske, Rev. Sci. Instrum. **60,** 1200 (1989).

[19] Machinable ceramic from Corning Glass Corporation (http://www.corning.com/specialtymaterials/products_capabilites/macor.aspx)

[20] Low temperature glue, STYCAST 2850FT, from Emerson and Cuming.

[21] Coarse positioner from Attocube Systems AG (model ANPz51).

[22] Piezo electric tube from EBL Products, Model EBL#2

[23] Cernox sensor from LakeShore Cryotronics Inc., USA.

[24] We found the two component silver epoxy EPO-TEK® E4110 from Epoxy Technology, Inc to have adequate mechanical strength for cleaving most single crystals.

[25] Ion source from Tectra GmbH, Frankfurt, Germany. Model: IonEtch sputter Ion Gun, Gen II

[26] Janis Research Company, USA. http://www.janis.com/

[27] Custom built SmartTable® with central hole from Newport Corporations, USA.

[28] Tinkham, M Introduction to Superconductivity (Dover Publications Inc., Mineola, New York, 2004).

[29] I. Giaever and K. Megerle, Phys. Rev. **122,** 1101 (1961)**.**

[30] R. V. Coleman, B. Giambattista, A. Johnson, W. W. McNairy, G. Slough, P. K. Hansma, and B. Drake, J. Vac. Sci. Technol. A 6, 338 (1988).

[31] M. Marz, G. Goll, and H. v. Löhneysen, Rev. Sci. Instrum. 81, 045102 (2010).

[32] See supplementary material at **[URL will be inserted by AIP]** for the variation of vortex lattice constant as a function of magnetic field.
16

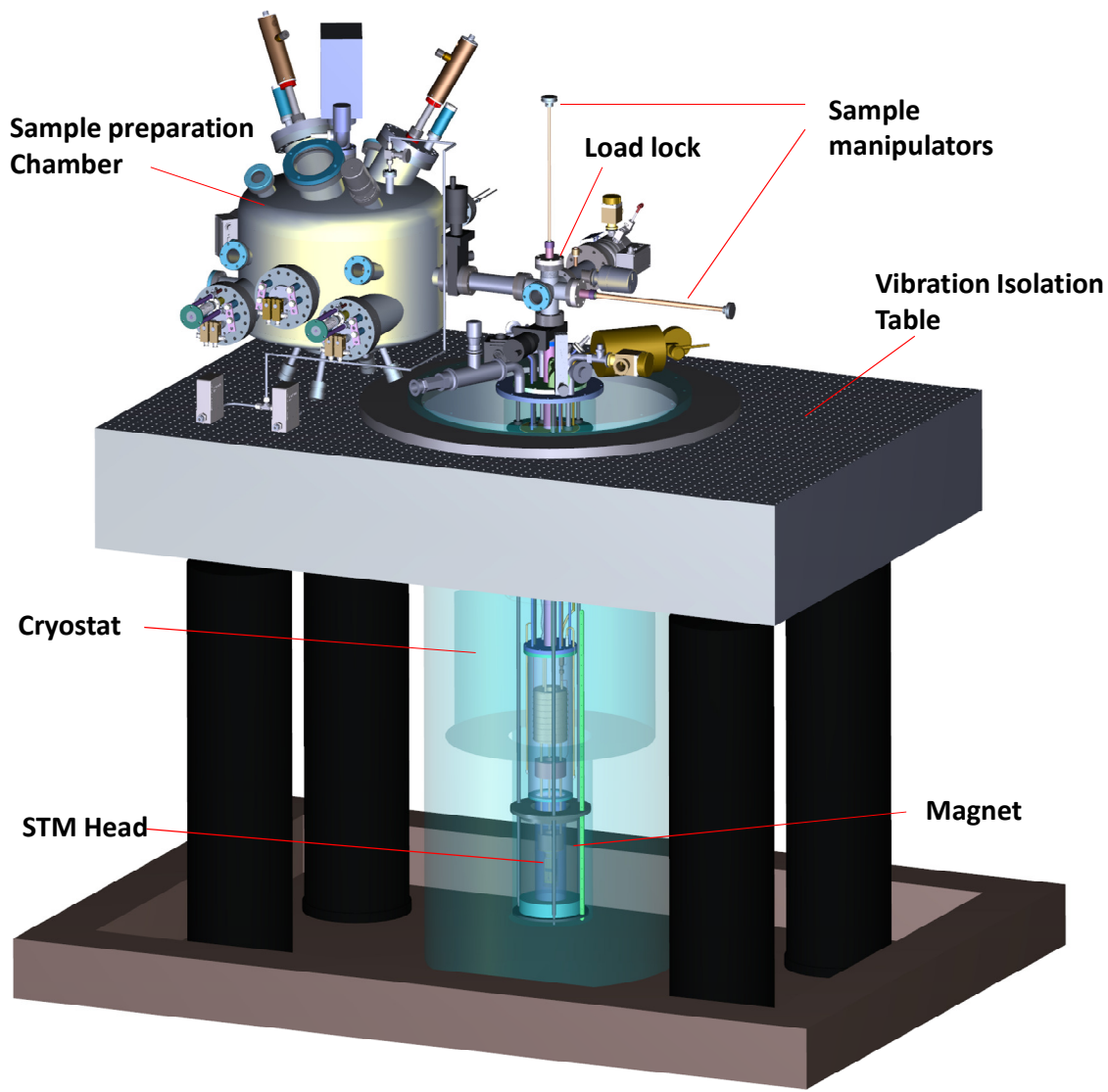

**Figure 1**. 3D view of the LT-STM assembly consisting of three primary sub-units: (i) The sample preparation chamber, (ii) the load lock chamber to transfer the sample from the deposition chamber to the STM and (iii) the 4He dewar with 9T magnet housing 3He cryostat on which the STM head is attached. The 4He dewar hangs from a specially designed vibration isolation table mounted on pneumatic legs. The Dewar, cryostat and magnet have been made semi-transparent to show the internal construction.

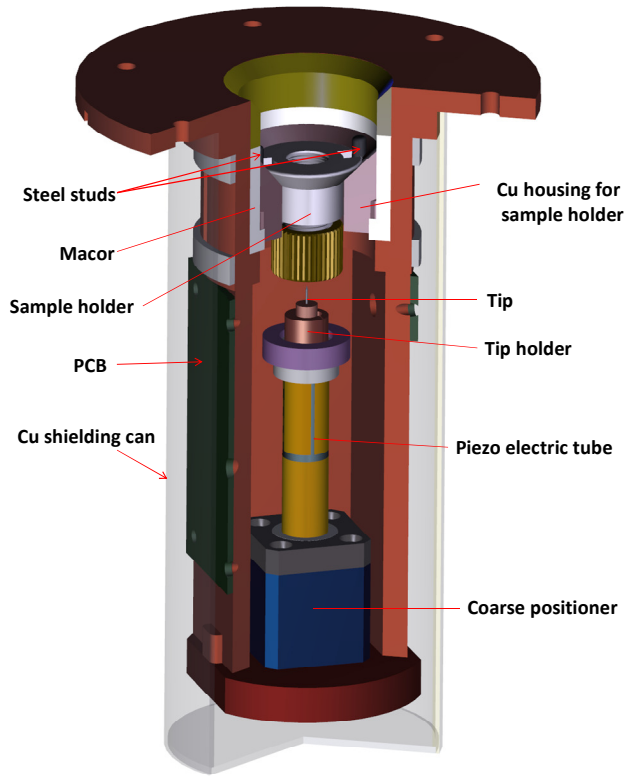

**Figure 2**. 3D view showing the construction of the STM head with the coarse positioner, piezoelectric scan-tube mounted, tip holder and sample holder with the sample facing down. The main body of the STM head is made of gold-plated copper.

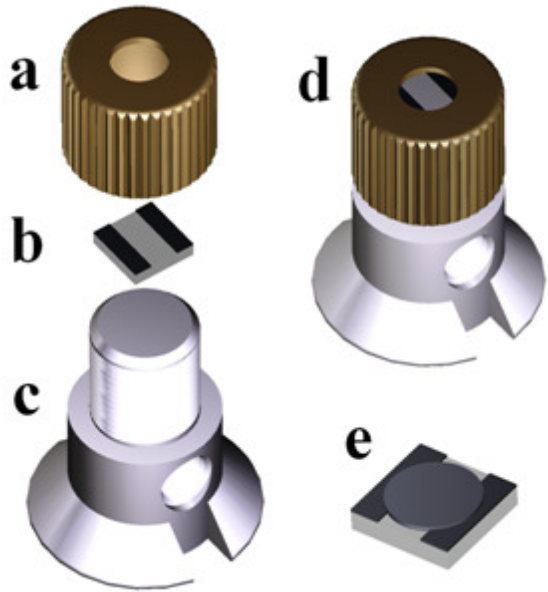

**Figure 3**. Design of sample holder (a) Molybdenum cap, (b) Substrate with strip deposited at the edge, (c) Molybdenum sample holder, (d) Sample holder assembly, showing substrate fastened with cap; (e) Resulting film on the substrate after the deposition.

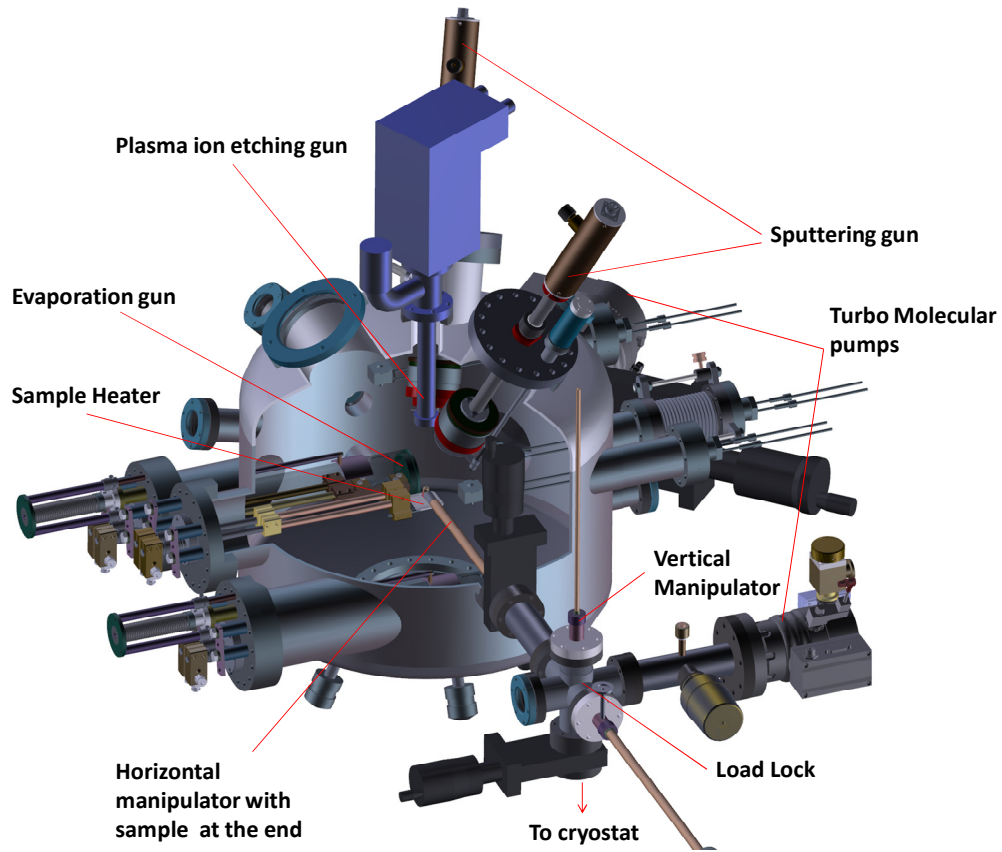

**Figure 4**. Schematic 3D view of the sample preparation chamber and load-lock cross. The deposition chamber incorporates two magnetron sputtering guns, a substrate heater for heating the substrate up to $800^0$ C, a plasma ion etching gun and two thermal evaporation sources. The substrate in inserted inside the deposition chamber using the horizontal manipulator.

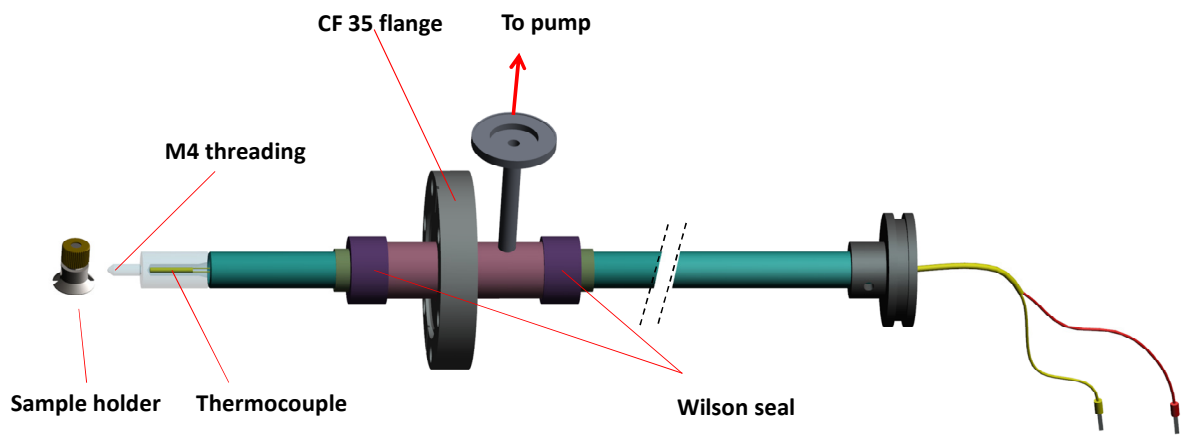

**Figure 5.** Design of the horizontal sample manipulator with in-built thermocouple for measuring the temperature during sample deposition. A differential pumping arrangement between two Wilson seals is used to remove any leaked gas during movement. The end of the manipulator is made transparent to show the position of the thermocouple. The vertical manipulator is similar in construction but does not have the thermocouple.

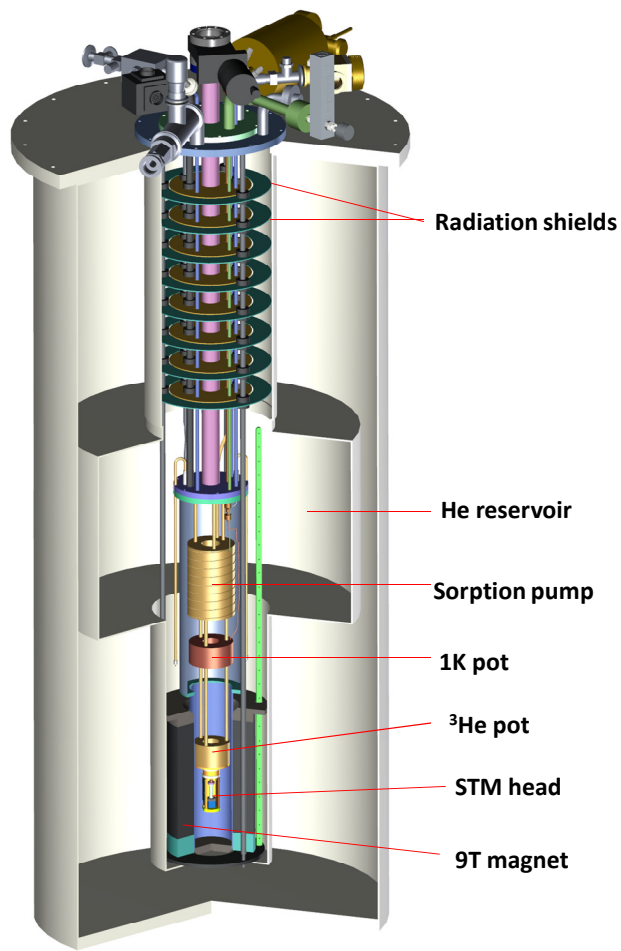

**Figure 6.** Schematic view of the $^3$He cryostat $^4$He dewar showing the $^4$He reservoir, the sorption pump, 1K pot, $^3$He pot and the STM head which is bolted below the $^3$He pot. The $^4$He Dewar has a capacity of 65 litres and a retention time of 5 days.

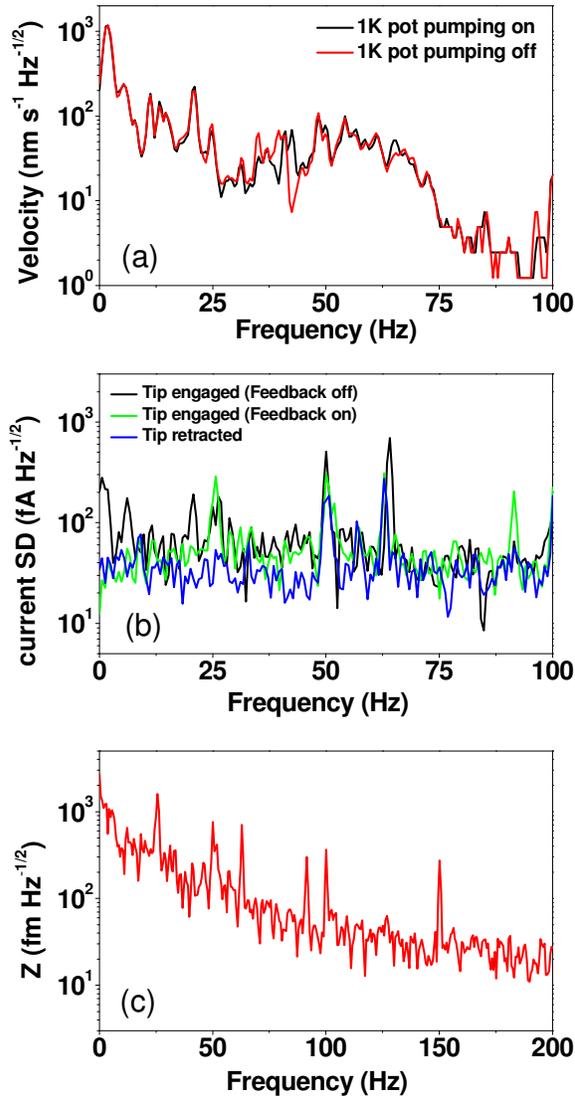

**Figure 7.** (a) Spectral density of the velocity vs. frequency on the top of the cryostat measured using an accelerometer. The spectral densities with and without the 1K pot pump on are nearly identical. (b) Spectral density of the tunneling current with the tip out of tunneling range, within tunneling range with feedback on and with feedback off. (c) Spectral density of Z height signal with feedback on. Measurements in (b) and (c) were performed at 350 mK on a $NbSe_2$ single crystal with tunneling current set to 50 pA and bias voltage to 20 mV.

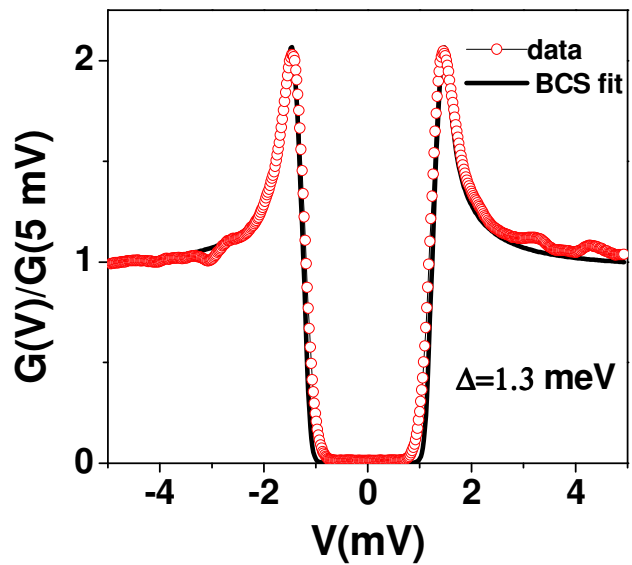

**Figure 8.** Tunneling spectroscopy on Pb single crystal acquired with Pt-Ir tip at 500mK along with BCS fit. The spectrum is averaged over 10 voltage sweeps at the same point. The spectroscopy set point before switching off the feedback was V = 6 mV, I = 500 pA, and the lock-in modulation voltage was 150 µV with frequency of 419.3 Hz.

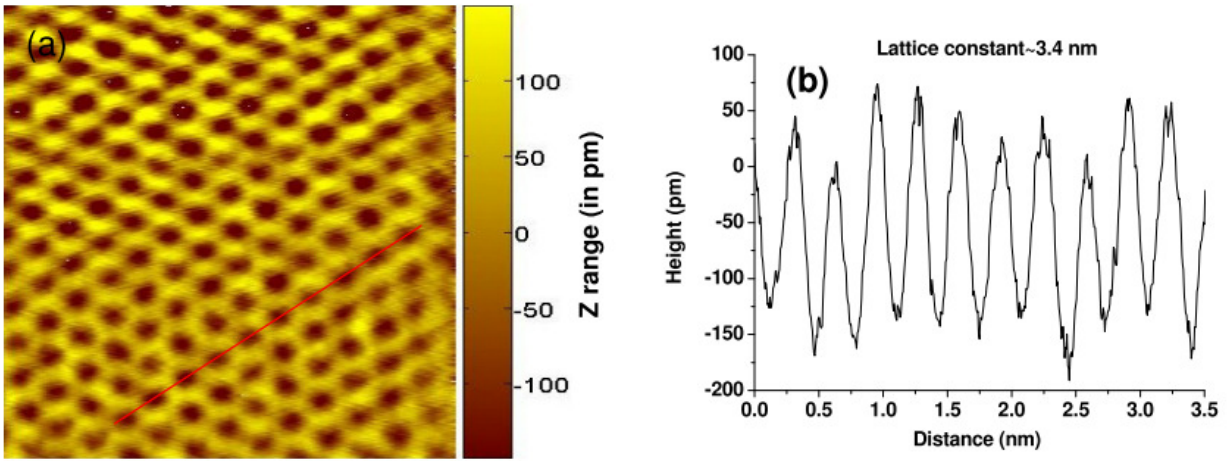

**Figure 9**. (a) Atomically resolved topographic image of NbSe$_2$ (4 × 4 nm) obtained in constant current mode. The tunneling current was set to 150 pA, the bias voltage to 20 mV and the scan speed was 13nm/s. (b) Line cut along the line shown in red in (a).

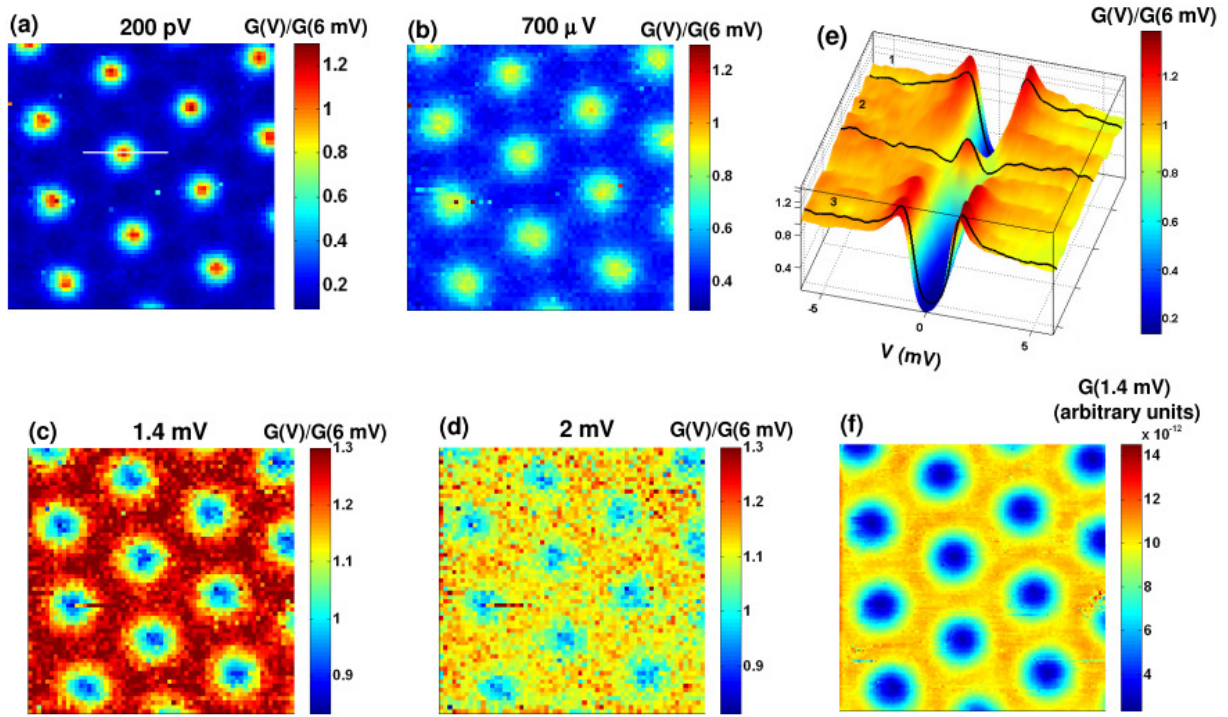

**Figure 10**. Vortex imaging on NbSe$_2$. (a)-(d) 64 × 64 conductance maps over 352 × 352 nm area at different voltages at 350 mK in an applied field of 200 mT. The maps are obtained from full spectroscopic scans from -6 mV to 6 mV at each pixel. (e) Line scan of the tunneling spectra along the white line marked in panel (a). Three spectra inside (2) and outside (1 & 3) vortex cores are highlighted in black. (f) High resolution conductance map acquired over the same area by scanning at fixed bias of V=1.4 mV; the tunneling current was set to 50 pA and modulation voltage was set to 150 μV with frequency of 2.3 kHz.

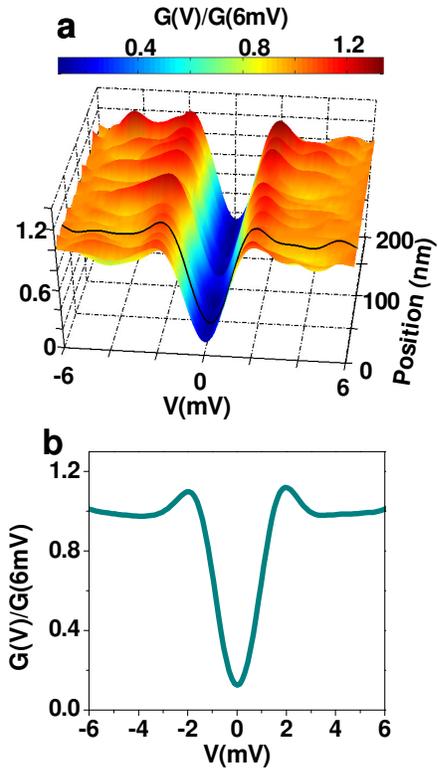

**Figure 11**. STS data for NbN thin film with $T_c$ = 6.4K. (a) Normalized tunneling spectra acquired along the line of length 200nm. (b) Average tunneling spectrum over 200 × 200 nm area. The modulation voltage was set to 150 µV with frequency of 2 KHz.

# A 350mK, 9T scanning tunneling microscope for the study of superconducting thin films and single crystals

Anand Kamlapure, Garima Saraswat, Somesh Chandra Ganguli, Vivas Bagwe and Pratap Raychaudhuri*

*Department of Condensed Matter Physics and Materials Science, Tata Institute of Fundamental Research, Homi Bhabha Rd., Colaba, Mumbai 400 005, India.*

and

Subash P. Pai

*Excel Instruments, 28, Sarvodaya Industrial Premises, Off Mahakali Caves Road, Andheri (East), Mumbai 400 093, India.*

**Vortex Lattice Constant as a function of magnetic field**

Figure 1s (a)-(h) show the images of the vortex lattice over 352 × 352 nm area, recorded at magnetic fields varying from 0.2 T to 2.8 T at 350 mK. For a hexagonal Abrikosov lattice, the nearest neighbour distance at magnetic field B is given by

$$a(B) = 1.075 \left(\frac{\Phi_0}{B}\right)^{\frac{1}{2}} \approx \frac{48.9}{\sqrt{B(in\ T)}}\ nm. \qquad (1)$$

In figure 1(g) we plot $a$ vs $B^{-1/2}$ obtained from the 8 images (points) along with expected variation from equation (1), showing excellent agreement with theory.

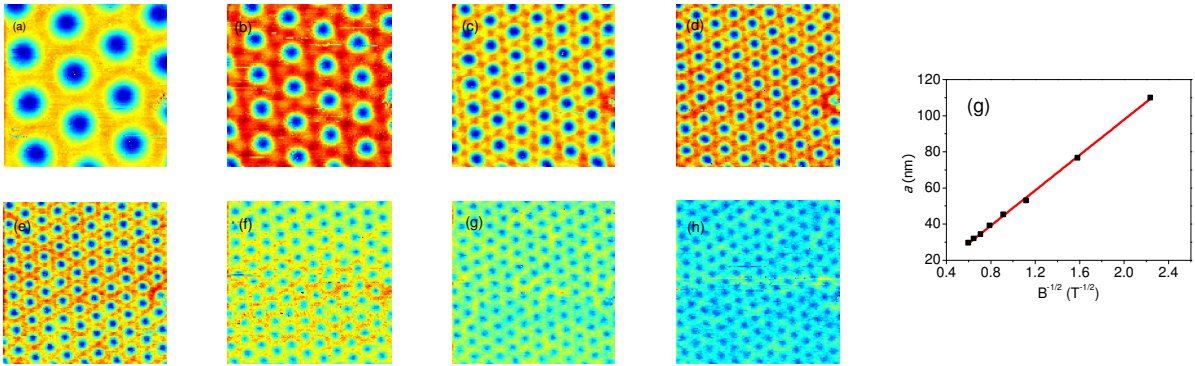

**Figure 1s.** Vortex lattice images on $NbSe_2$ over 352 × 352 nm area at 350 mK acquired at magnetic fields of (a) 0.2 T, (b) 0.4 T, (c) 0.8 T, (d) 1.2 T, (e) 1.6 T, (f) 2 T, (g) 2.4 T and (h) 2.8 T. The images are acquired by measuring $dI/dV$ at a fixed bias voltage of 1.4 mV. The tunnelling current was set to 50 pA and modulation voltage was 150 µV with frequency of 2.3 kHz. (g) Nearest neighbour distance of the vortices, $a$, as a function of $B^{-1/2}$ (black points) along with the expected variation from eq. (1) (red line).